\documentclass[twocolumn,floatfix,prb,superscriptaddress]{revtex4-1}

\usepackage{amsmath,amssymb,bm,graphicx,url}
\usepackage{subfigure}
\usepackage{graphicx}
\usepackage{psfrag}
\usepackage{amssymb}
\usepackage[dvipsnames]{xcolor}
\usepackage{bbold}

\newcommand{\Eq}[1]{equation \eqref{#1}}
\newcommand{\Eqs}[2]{equations~\eqref{#1} and \eqref{#2}}

\newcommand{\Fig}[1]{Fig.~\ref{#1}}
\newcommand{\Tab}[1]{Table~\ref{#1}}
\newcommand{\Lio}{\mathcal{L}}

\newcommand{\be}{\begin{equation}}
\newcommand{\ee}{\end{equation}}
\newcommand{\bea}{\begin{eqnarray}}
\newcommand{\eea}{\end{eqnarray}}

\newcommand{\bra}[1]{\left\langle \, #1 \right|}
\newcommand{\ket}[1]{\left| #1  \right\rangle}

\newcommand{\expec}[1]{\left\langle #1 \right\rangle} 
\newcommand{\comm}[2]{\left[ #1, #2 \right]}
\newcommand{\lind}[1]{\mathcal{D}\left[#1\right]}
\newcommand{\coup}[2]{\mathcal{C}\left[#1, #2 \right]}
\newcommand{\meas}[1]{\mathcal{M}\left[#1\right]}

\newcommand*{\citein}[1]{%
  \begingroup
    \romannumeral-`\x 
    \setcitestyle{square,numbers}%
    Ref.\cite{#1}%
  \endgroup   
}

\newcommand{\sq}[1]{\left[ {#1} \right]}
\newcommand{\tr}[1]{{\textrm {Tr}}\sq{#1}}

\newcommand{\op}[2]{\ket{#1}\bra{#2}}

\newcommand{\dg}{^\dagger}

\newcommand{\beq}{\begin{equation}}
\newcommand{\eeq}{\end{equation}}
\newcommand{\bqa}{\begin{eqnarray}}
\newcommand{\eqa}{\end{eqnarray}}
\newcommand{\nn}{\nonumber}

\newcommand{\Tr}{\mbox{Tr}}

\begin{document}

\title{Quantum nondemolition detection of a propagating microwave photon}

\author{Sankar R. Sathyamoorthy}
\affiliation{Microtechnology and Nanoscience, Chalmers University of Technology, S-41296 Gothenburg, Sweden}
\author{L. Tornberg}
\email[Email: ]{lars.tornberg@chalmers.se}
\affiliation{Microtechnology and Nanoscience, Chalmers University of Technology, S-41296 Gothenburg, Sweden}
\author{Anton F. Kockum}
\affiliation{Microtechnology and Nanoscience, Chalmers University of Technology, S-41296 Gothenburg, Sweden}
\author{Ben Q. Baragiola}
\affiliation{Center for Quantum Information and Control, University of New Mexico, Albuquerque, NM 87131-0001, USA}
\author{Joshua Combes}
\affiliation{Center for Quantum Information and Control, University of New Mexico, Albuquerque, NM 87131-0001, USA}
\author{C.M. Wilson}
\affiliation{Microtechnology and Nanoscience, Chalmers University of Technology, S-41296 Gothenburg, Sweden}
\affiliation{Institute for Quantum Computing and Electrical and Computer Engineering Department, University of Waterloo, Waterloo N2L 3G1, Canada}
\author{Thomas M. Stace}
\affiliation{Centre for Engineered Quantum Systems, School of Physical Sciences, University of Queensland,
Saint Lucia, Queensland 4072, Australia}
\author{G. Johansson}
\affiliation{Microtechnology and Nanoscience, Chalmers University of Technology, S-41296 Gothenburg, Sweden}
\begin{abstract}
The ability to nondestructively detect the presence of a single, traveling photon has been a long-standing goal in optics, with  applications in quantum information and measurement. Realising such a detector is complicated by the fact that photon-photon interactions are typically very weak. At microwave frequencies, very strong \emph{effective} photon-photon interactions in a waveguide have recently been demonstrated. Here we show how this type of interaction can be used to realize a \emph{quantum nondemolition measurement} of a single propagating microwave photon.  The scheme we propose uses a chain of  solid-state 3-level systems (transmons), cascaded through circulators which suppress photon backscattering.  Our theoretical analysis shows that microwave-photon detection with fidelity around 90\% can be realized with existing technologies.
\end{abstract}

\maketitle

Quantum mechanics tells us that a measurement perturbs the state of a quantum system. In the most extreme case, this leads to the destruction of the measured quantum system. By  coupling the system to a quantum probe, a quantum nondemolition \cite{BraginskyKhalili} (QND) measurement can be realized, where the system is not destroyed by the measurement. Such a property is crucial for quantum error correction \cite{SteanePRL96}, state-preparation by measurement \cite{RuskovPRB2003, BishopNJP2009} and one-way quantum computing \cite{RaussendorfPRL2001}.   For microwave frequencies, detection of confined photons in high-Q cavities has been proposed and experimentally demonstrated by several groups \cite{SchusterNature2007, GuerlinNature2007, WangPRL2008, WangPRL2008, JohnsonNature2010}. They all exploit the strong interaction between photons and atoms (real and artificial) on the single photon level. Detection schemes for traveling photons have also been suggested \cite{RomeroPRL2009, ChenPRL2011, PeropadrePRA2011}, but in those proposals the photon is absorbed by the detector and the measurement is therefore destructive. Proposals for detecting itinerant photons using coupled cavities have also been suggested, but they are limited by the trade-off between interaction strength and signal loss due to reflection \cite{HelmerPRA2009}.  Other schemes based on the interaction of $\Lambda$-type atomic level structures have been suggested, but the absence of such atomic level structures in solid-state systems make them unsuited to the microwave regime \cite{WitthautEPL2012,MunroPRA2005, GreentreeNJP2009}.

\begin{figure}
  \includegraphics[width=\columnwidth]{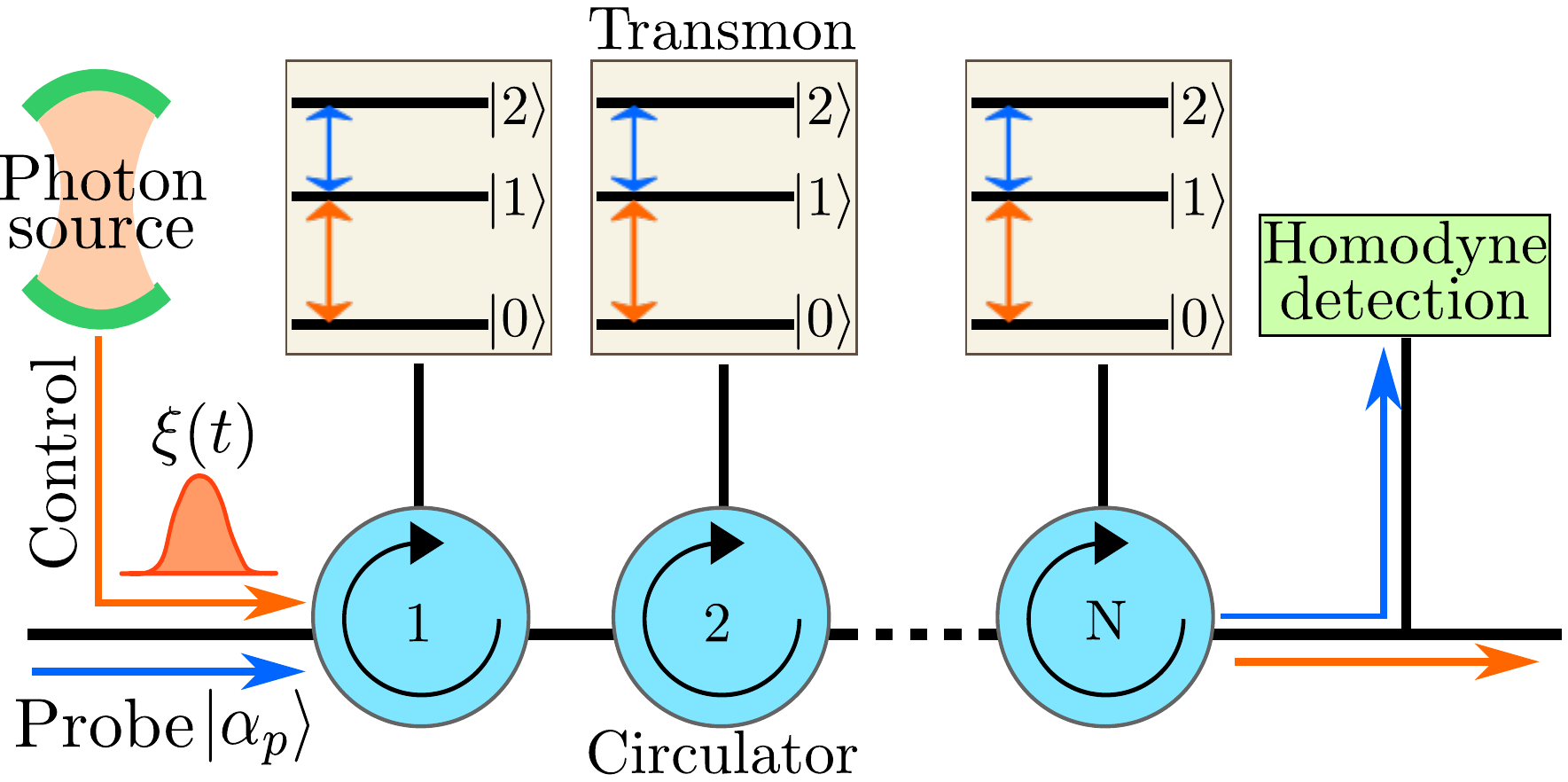} 
     \caption{
     A chain of $N$ transmons cascaded from microwave circulators interacts with control and probe fields, which are close to resonance with the 0-1 and 1-2 transition respectively. In the absence of a control photon, the chain is transparent to the probe. A control photon with temporal profile $\xi(t)$ drives each transmon consecutively, which then displaces the probe field, which is detected by homodyne measurement.     } 
     \label{fig:multi_mirror_setup}
 \end{figure}
 
\begin{figure*}
\centering
\includegraphics[width=17cm]{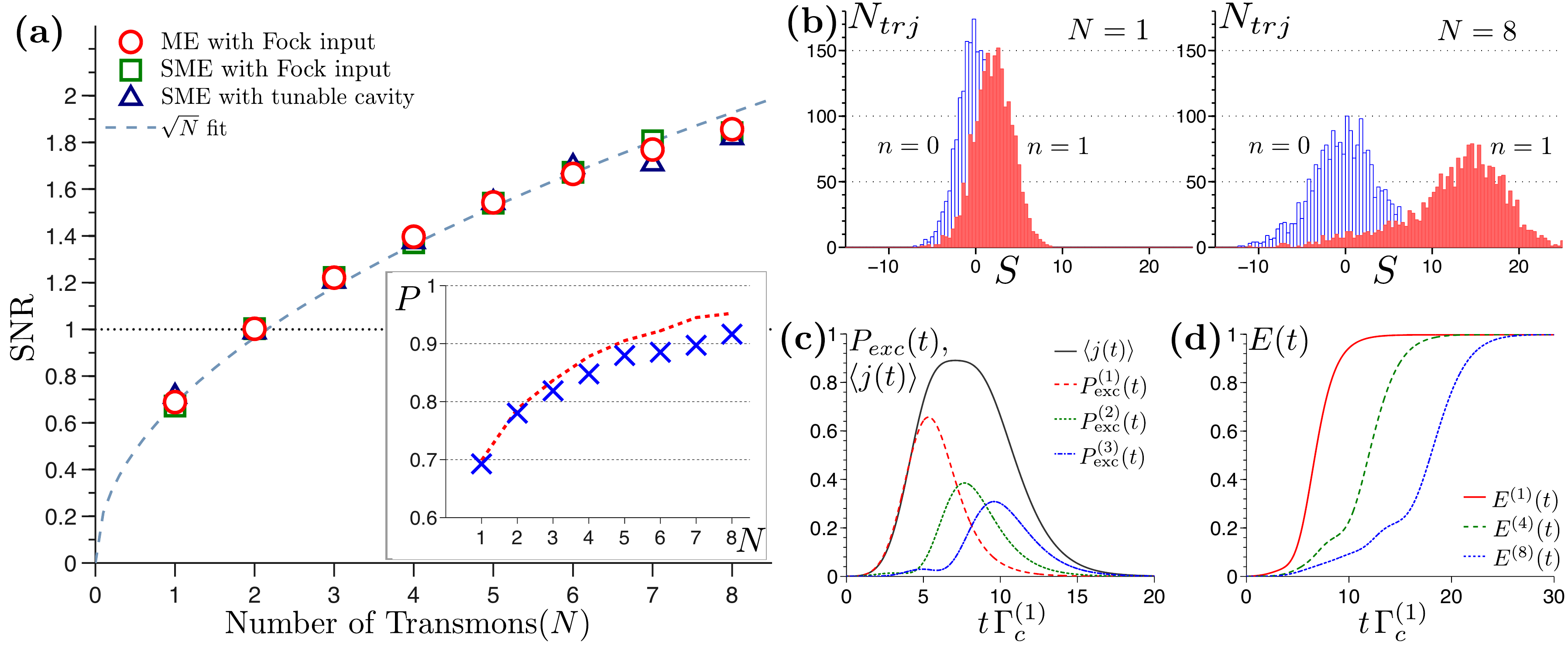}
\caption[]{ Signal-to-noise ratio (SNR) and detector dynamics. (\textbf a) SNR calculated using the master equation with Fock state input [SM] [red circles], stochastic master equation with a Fock state input [green squares] and stochastic master equation with a tunable cavity as photon source [blue triangles]. The dashed line shows a $\sqrt{N}$-fit assuming that each transmon contributes equally and independently to the signal. (inset to \textbf{a})  Detection fidelity, $P$, of correctly inferring the control photon number, \Eq{F}.  The crosses are results of Monte-Carle sampling from the stochastic master equation;  the dashed line is inferred from the SNR, assuming a normal distribution.  (\textbf b) Histogram of signal with 0 [blue] and 1 [red] control photons. For $N=1$ transmon the distributions overlap significantly (SNR=0.7); for $N=8$ the distributions are well-resolved (SNR=1.85).   (\textbf c) Transmon excitation, $P_{\textrm{exc}}$, and average homodyne current, $\langle j(t)\rangle$,  over time for $N=3$. (\textbf d) Integrated output photon flux for $N=1,4$ and 8, showing unity transmission.  Transmon parameter values from \Tab{tab:parameters};  homodyne  local oscillator phase $\phi=\pi/2$;  probe integration window $4<t<8+1.5(N-1)$; a value which is found by numerical optimization to limit the amount of added noise to the signal. 
All quantities are in units of $\Gamma_c^{(1)} = \Gamma_{01}^{(1)}$.}
\label{fig:SNR}
\end{figure*}

Here, we present a   scheme to detect a propagating microwave photon in an open waveguide.  At its heart is the strong \emph{effective} nonlinear interaction between microwave fields induced by an artificial atom to which they are coupled. A single photon in the \emph{control} field induces a detectable displacement in the state of a \emph{probe} field, which is initially in a coherent state. The control field is not absorbed, making the protocol QND.  The protocol may be operated either synchronously (in which the control photons arrive within specified temporal windows), or asynchronously \cite{ReisererScience2013}.

\Fig{fig:multi_mirror_setup} illustrates the scheme.  The effective nonlinear interaction between the control photon and the probe field is realized by $N$ noninteracting artificial atoms (\emph{transmon} devices \cite{KochPRA2007}) coupled to the transmission line.  Transmons are particularly attractive in light of recent work demonstrating strong atom-field coupling in the single-photon regime in open waveguides \cite{IoChunPRL2013}. We treat the atoms as anharmonic three-level ladder systems with energy eigenstates $\ket{0}, \ket{1}$ and $\ket{2}$ designed such that the control  and probe fields are close to resonance with the $\ket{0}\leftrightarrow\ket{1}$ and $\ket1 \leftrightarrow\ket2$ transition, respectively.  In the absence of a control photon, the system is completely transparent to the probe. A single control photon will sequentially excite the chain of transmons, displacing the state of the probe field  \cite{IoChunPRL2013}, which we show can be detected by  homodyne measurement of the probe.

While a single transmon in a waveguide induces a large photon-photon nonlinearity, the induced displacement of the probe field is nevertheless insufficient to yield a useful detection protocol \cite{BixuanPRL2013}. Furthermore, Kramers-Kronig relations impliy there will be substantial back-reflection of the control photon from a strongly coupled transmon, and it was shown that this precludes cascading additional transmons in an attempt to boost the probe displacement \cite{BixuanPRL2013}.  In order to evade this issue, we propose cascading transmons from stub waveguides attached to a chain of circulators, shown in \Fig{fig:multi_mirror_setup}.
In this geometry, the circulators suppress  back-scattering \cite{StannigelNJP2011}, and the fields propagate unidirectionally along the waveguide.  Thus, multiple transmons can interact with the control photon, which is fully transmitted to the output.


To provide a quantitative analysis of this proposal, we start by defining the transmon Hamiltonian ($\hbar = 1$)
\be
H = -\omega_{01} \ket{0}\bra{0} + \omega_{12} \ket{2}\bra{2}.
\ee
We couple the transmons within a cascaded one-way channel which we treat using the input-output formalism \cite{CarmichaelPRL1993,GardinerPRL1993,GoughCommMathPhys2009}.  The final output probe field of the chain yields information about the state of the control field, which we quantify through a suitably defined signal-to-noise ratio (SNR), and through a fidelity measure, $P$, the probability to infer the correct photon number in the control field.


We take two equivalent approaches to model the photon source \cite{BixuanPRL2013}. In the first, we invoke a fictitious source cavity with resonance frequency $\omega_c$ and annihilation operator, $a$, to generate the control photon.  In the rotating frame defined by the unitary transformation $U(t) = \exp(it( \omega_c(\ket{0}\bra{0} +   a^\dagger a) + \omega_p\ket{2}\bra{2}))$, where $\omega_{p/c}$ is the probe/control frequency, the dynamics of the cascaded system, including the control photon and back-action of the homodyne measurement is given by the stochastic master equation (See Supplemental Material [SM])
\begin{eqnarray}
d\rho &=& -i\comm{H_\text{eff}}{\rho} dt  + \sqrt{\eta} \meas{\Lambda_{12}}{\rho} \; dW(t) \nonumber\\
&+&\Bigg[\kappa \,\lind{a}- \sqrt{\kappa}\, \coup{a}{\Lambda_{01}}+ \sum_{j=1}^N\bigg(\lind{L_{01}^{(j)}} + \lind{L_{12}^{(j)}}   \nonumber \\
&-& \sum_{k=j+1}^N \left(\coup{L_{01}^{(j)}}{L_{01}^{(k)}} + \coup{L_{12}^{(j)}}{L_{12}^{(k)}}   \right)\bigg) 
\Bigg] \rho \,dt,  \label{Eq_SMECavity}
\end{eqnarray}
where $H_\text{eff} = \sum_{k=1}^{N} H^{(k)}$ is the effective Hamiltonian with the single transmon Hamiltonian given by
$
{H}^{(k)}= -\Delta^{(k)}_c\ket{0}\bra{0}^{(k)}+\Delta^{(k)}_p\ket{2}\bra{2}^{(k)} + \Omega_p(L^{(k)}_{12} + L^{(k)}_{21} ),
$
 $\Delta_c^{(k)} = \omega_{10}^{(k)} - \omega_c$ and $ \Delta_p^{(k)} = \omega_{21}^{(k)}-\omega_p$ are the detunings between control and probe fields and the  transition frequencies of transmon $k$, and $\Omega_p$ is the probe amplitude. We have defined $\lind{c} \rho = c\rho c^{\dagger }-\frac{1}{2}c^{\dagger }c\rho -\frac{1}{2} \rho c^{\dagger }c$  \cite{LindbladCommMathPhys1976}, $L_{ij}^{(k)}= \sqrt{\Gamma_{ij}^{(k)}} \ket{i}\bra{j}^{(k)}$, and $\Lambda_{ij}=\sum_{k=1}^{N} L_{ij}^{(k)}$, where $\Gamma_{ij}^{(k)} = \Gamma_{ji}^{(k)}$ are transmon decay rates. We allow  different couplings between each transmon and the waveguide, which we numerically optimize to achieve high SNR. 
 The cascaded, field-induced interaction between the transmons in the chain is described by the superoperator
$\coup{c_1}{c_2} \rho = [c_2^{\dagger }, c_1\rho ]+[\rho\,c_1^{\dagger },c_2]$ 
and $\meas{c}\rho = (e^{i\phi}c\,\rho + e^{-i\phi}\rho \,c^{\dagger })-\expec{e^{i\phi}c+ e^{-i\phi} c ^{\dagger }}\rho$ is the measurement superoperator describing the back-action of the homodyne measurement with local oscillator phase $\phi$ and efficiency $\eta$ \cite{WisemanPRA1993}. $dW(t)$ is a Wiener increment satisfying $E[dW(t)]=0$ and $E[dW(t)^2]=dt$.
The control photon envelope $\xi(t)$ is set by the time-dependent cavity damping rate $\kappa(t) = \xi(t)/({\int_t^\infty |\xi(s)|^2ds})^{1/2}$ \cite{GoughPRA2012}. 
Although the cavity considered here is notionally fictitious and serves merely as the photon source, we note that photon generation with cavities with/without tunable couplings have been demonstrated recently \cite{YinPRL2013, LangNatPhys2013, PechalArXiv2013}. 

In the second approach, we adopt the Fock state master equation formalism, in which the control photon envelope appears explicitly \cite{BaragiolaPRA2012}[SM]. In either of the approaches, the transmon dynamics are \emph{not} adiabatically eliminated, but included explicitly (within the rotating wave approximation). Hence we do not rely on any effective cross-Kerr type Hamiltonian to mediate photon-photon interactions \cite{ImotoPRA1985, MunroPRA2005}.

\begin{figure}
  \includegraphics[width= \columnwidth]{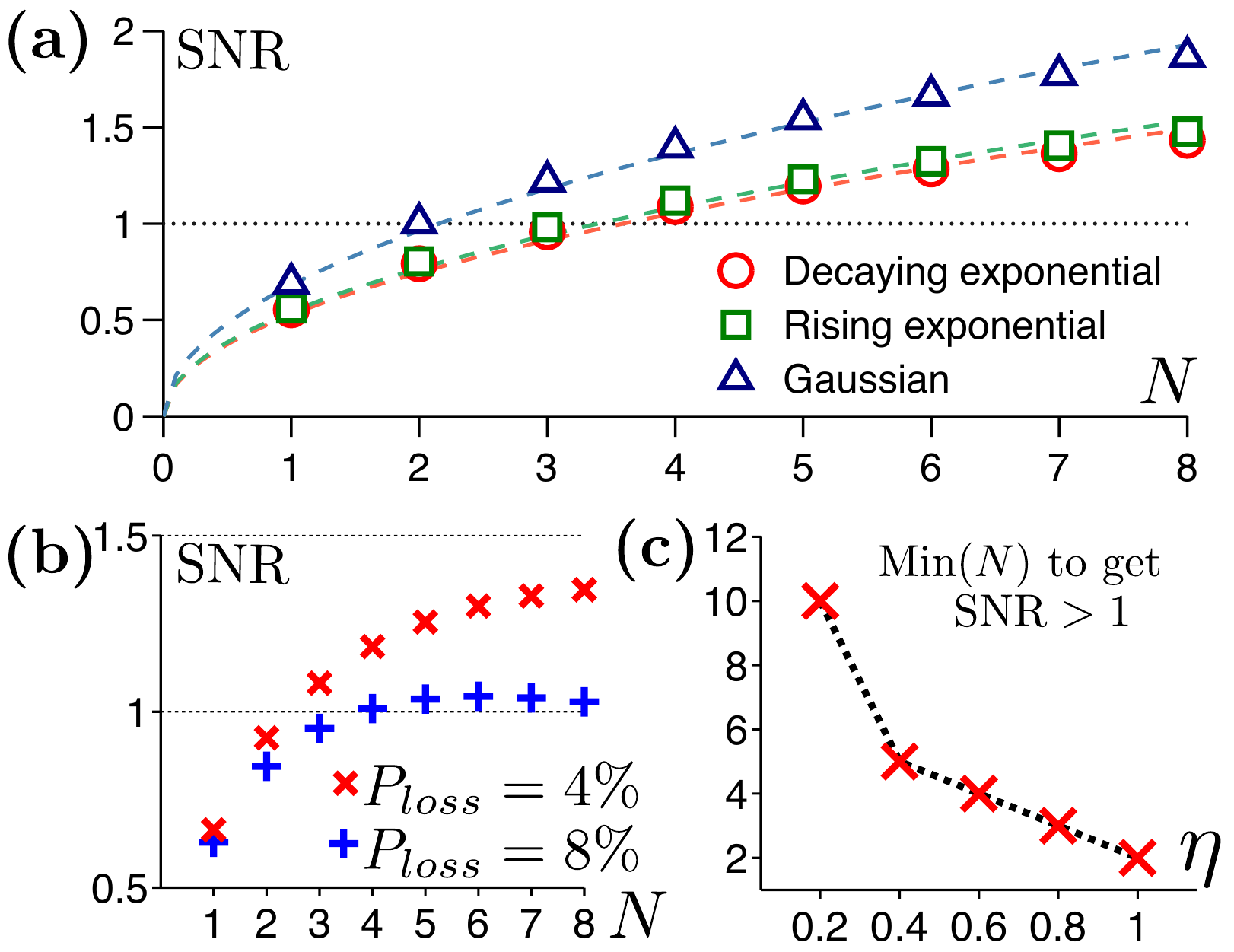} 
     \caption{
    (\textbf a) SNR with different input photon shapes. 
    The parameters of the transmons and the probe field are optimized for each of the pulse shape (see Table I). The dashed lines are fitted using $\chi \sqrt{N}$; the best-fit values of $\chi$ are in Table I. (\textbf b) Effect of losses in circulators on the SNR for a Gaussian input photon for two realistic choices of power loss, $P_{loss}$, in each circulator. (\textbf c) Number of transmons that would be required to achieve SNR $\ge1$ for a Gaussian input photon, for detector efficiency $\eta$. }  \label{fig:exp}
 \end{figure}

In either numerical simulations, or in an actual experimental implementation of this scheme, the stochastic output from the homodyne measurement of the probe field is $
j_n(t) dt = \sqrt{\eta} \expec{e^{i\phi}\Lambda_{12}+e^{-i \phi}\Lambda_{21}} dt + dW(t)
$, where $n$ is the number of photons in the control field. 
The ultimate objective is to derive a signal, $S_n$, encoding the state of the control field from the homodyne current $j_n$. The most straightforward approach is to average $j_n$ over a suitably optimised temporal window $S_n = \int_{t_{i}}^{t_{f}} j_n(t) dt$, where $t_m = t_f - t_i$ is the measurement window.  In the absence of a control photon,  $j_n$ consists only of the quantum noise associated with the probe field, and so  $\mathrm{E}[S_0]=0$, and the variance is $\mathrm{Var}[S_0] = t_m$. In the presence of a control photon, the probe field is displaced in phase space, and the homodyne current acquires a time-dependent component (pictured in Fig.\ 2c), so $\mathrm E[S_1]\neq0$.  Thus, we define the signal-to-noise ratio 
\be
\mathrm{SNR} = {\mathrm{E}[S_1]}/{\sqrt{\mathrm{Var}[S_1] + t_m}}.
\ee 
We note that there are more optimal ways to define the signal, $S_n$, using linear or nonlinear filters on $j_n$ \cite{GambettaPRA2007}. We find that optimal linear filters improve the SNR modestly [SM], but for brevity and simplicity, the results in this Letter use the simple definitions above.

\begin{table*}[]
\centering
\begin{tabular}{|c|c|c|c|c|c|c|c|c|c|c|c|c|c|}
\hline 
Photon Shape &$\xi(t)$ &$\Gamma_{ph}$ & $T_{ph}$ & $\Gamma_c^{(1)}$ & $\Gamma_c^{(2)}$ & $\Gamma_c^{(3)}$ & $\Gamma_c^{(4)}$ & $\Gamma_c^{(5)}$ & $\Gamma_c^{(6)}$ & $\Gamma_c^{(7)}$ & $\Gamma_c^{(8)}$ & $\Omega_p^2$ & $\chi$   \\ 
\hline 
Gaussian & $\left(\Gamma_{ph}^2/(2\pi) \right)^{1/4} \exp\left( -\Gamma_{ph}^2(t-T_{ph})^2/4\right)$ &0.8 & 4 & 1.0 & 1.9 & 2.2 & 2.5 & 2.4 & 2.5 & 2.7 & 3.2 & 0.12 & 0.6813 \\ 
\hline 

Decaying exponential & $\Theta(T_{ph}-t)\sqrt{\Gamma_{ph}} \exp\left(-\Gamma_{ph} t/2\right)$ & 0.5 & 4 & 1.0 &1.6 &2.1 &2.5 &2.6 &2.9 &3.5  & 3.8& 0.16 & 0.5272 \\ 
\hline 
Rising exponential & $\Theta(t-T_{ph})\sqrt{\Gamma_{ph}} \exp\left(\Gamma_{ph} t/2\right)$ &0.5 & 12 &1.0 &1.9 &2.3 &2.6 &3.0 &3.3 &3.5  & 3.8 & 0.16 & 0.5424\\ 
\hline 
\end{tabular}
\caption{Parameter values for which the SNR is plotted in Fig.\ref{fig:SNR} and Fig.\ref{fig:exp}. All of the rates are in units of $\Gamma_c^{(1)}$. The values of other parameters are $\Gamma_{p}^i = 2 \Gamma_{c}^i$, $\Delta_c^i = 0$, $\Delta_p^i = 0$. 2000 trajectories are used in the simulation for each control input state.}
\label{tab:parameters}
\end{table*}

The mean of the homodyne current, $\langle j(t)\rangle$,  is given by the unconditional dynamics of \Eq{Eq_SMECavity}. Using the Quantum Regression Theorem, it is  possible to  calculate the variance from the unconditional dynamics and one does not in principle need to solve the full stochastic evolution if interested in the signal-to-noise ratio as written above \cite{QuantMeasAndContrl}. However, since the SNR does not contain any information about the signal distribution, a quantitative assessment of the measurement fidelity  can only be done by Monte-Carlo sampling the output of the full stochastic master equation of \Eq{Eq_SMECavity}. We define the measurement fidelity, $P$, to be the probability of correctly inferring the state of the control field,
\be
\label{F}
P = P(S < S_0^T|0)P(n=0) + P( S > S_1^T|1)P(n=1)
\ee 
where $P(S \leq S_0^T|n)$ is the conditional probability that $S$ is less than  some predefined threshold, $S_n^T$, given the control field has $n$ photons.  For the purposes of this Letter, we assume the priors $P(0)=P(1)=1/2$.  

Figure \ref{fig:SNR}a is the main quantitative result of this Letter, showing SNR as a function of the number of transmons, $N$, assuming a Gaussian control-photon envelope $\xi(t)$
(see \Tab{tab:parameters}). For a single transmon the SNR is around $0.7$, consistent with the  results in \cite{BixuanPRL2013}. For $N=2$ transmons the SNR reaches unity and grows monotonically with $N$.   \Fig{fig:SNR}b show the corresponding histograms of $S_{0,1}$ for $N = 1$ and $N = 8$.  Clearly, the peaks are not resolved for $N=1$, but they are well resolved for $N=8$, consistent with improvement in SNR. For the latter, there is a notable skewness and increase in the width of the distribution for the 1-photon case. Although this seems to indicate a slight memory effect, we can fit the SNR data in \Fig{fig:SNR}a to a simple $\sqrt{N}$ dependence,  which is what we expect if each transmon contributed equally and independently to the signal.  The crosses in the inset of \Fig{fig:SNR}a shows the probability of correctly inferring the input photon number (using \Eq{F}) assuming a common threshold, $S_0^T=S_1^T$. The dashed line is  a theoretical fit estimated from the SNR assuming Gaussian distributions, which slightly overestimates the fidelity, on account of the non-Gaussian skewness in the histograms.  

If a definitive measurement result is not required, the value of $P$ can be  improved by choosing separate thresholds, $S_0^T<S_1^T$, such that the data falling between the thresholds are rejected as inconclusive. 
Our analysis shows that with 8 transmons, a value of P = 0.95 (P = 0.90) can be reached with $15\%$ ($0\%$) of the measurement records being rejected. These numbers are above or comparable to the measurement efficiencies quoted in \cite{ RomeroPRL2009, ChenPRL2011, PeropadrePRA2011} and we stress the fact that the detection schemes proposed there were destructive. 

The detector dynamics is shown in \Fig{fig:SNR}c, showing the excitation probability of each transmon as function of time for $N=3$ along with the average homodyne signal, $\langle j(t) \rangle$. As the excitation is relayed through the chain, each transmon contributes to $S$,  overwhelming the quantum noise in the probe. The results in \Fig{fig:SNR} were obtained for an optimised set of parameter values by numerically tuning $\Delta_c$, $\Delta_p$, $\Gamma_c = \Gamma_{01}$ and $\Gamma_p =\Gamma_{12}$ independently for each transmon. We emphasize that this is quite robust: we can get $\rm{SNR}>1$, for a wide range of parameters [SM].  \Fig{fig:SNR}d shows the integrated control photon flux [SM], which asymptotes to unity, indicating that the detector is transparent to the control field, making the protocol QND, albeit with some distortion in the control photon envelope.  At the expense of a slightly diminished fidelity, $P$, we can choose the transmon coupling strengths to preserve the control photon envelope  [SM].  \Fig{fig:exp}a shows the SNR for non-Gaussian control-photon envelopes, $\xi(t)$, which are quantitatively somewhat worse than than the Gaussian case, but not qualitatively so.

Circulators underpin the operation of this proposal. In practice, commercial circulators suffer both insertion loss and limited isolation. Both of these effects can be modeled as losses appearing at the input to the circulator: insertion loss is obvious, while imperfect isolation manifests as backscattering or non-efficient interaction, which we treat as a loss process [SM].  We simulate both of these imperfections by interleaving fictitious beam-splitters at the input to each circulator  [SM]. \Fig{fig:exp}b  shows the SNR as a function of $N$ for $4\%$ and $8\%$ power loss . Since more power is lost with larger $N$, we see that the previously monotonically increasing SNR now acquires a maximum value. Importantly, we see that it is still possible to get SNR $>1$ with realistic numbers ($5\%- 10\%$) \cite{circulatorData} and we expect this to be an even smaller issue if on-chip circulators \cite{KochPRA2010} are realized.

\Fig{fig:exp}c shows the performance of the proposed detector when the efficiency of homodyne detection is less than 100\%.  Reassuringly, the overhead caused by an inefficient detector is not too high. 
With current state of the art amplifiers \cite{MalletPRL2011}, this scheme should be able to detect photons with a moderate number of transmons. The proposal is not impacted significantly by dephasing [SM]. 

In principle, the probe beam can be left on, since the detector is transparent to the probe when no control photons are present.  In our analysis we assume that the \emph{a priori} probability of a photon in a given temporal window is 50\%. This is consistent with applications where the possible arrival time of the photon is known \emph{a priori}.  However the detector may be operated in an asynchronous running mode (in which the \emph{a priori} control photon probability in a given temporal window is not 50\%), where the integration window of temporal width $t_m$ slides over the entire duration of the homodyne measurement, and photon arrivals will be marked by a peak in this moving average. 

In conclusion, we have proposed and analyzed a microwave photon detector consisting of a chain of transmon qubits. We show that a modest number of transmons connected by circulators is enough to achieve SNR$>1$, and a measurement fidelity $>90\%$. We anticipate that analogous protocols could be used to detect other itinerant bosonic particles e.g. phonons \cite{GustavssonNaturePhysics2012} and, together with postselection, be used for probabilistic generation of nonclassical states of the bosonic field.

\section{Acknowledgments}
We thank Per Delsing, Bixuan Fan, Io Chun Hoi and Gerard Milburn for fruitful discussions. We acknowledge financial support from the Swedish Research Council, the Wallenberg Foundation, STINT and the EU through the projects SOLID and PROMISCE. TMS is funded by the ARC Research Fellowship program and the ARC Centre of Excellence in Engineered Quantum Systems. BQB and JC acknowledge support from NSF Grant Nos. PHY-1212445 and PHY-1005540, AFOSR Grant No. Y600242, and ONR Grant No. N00014-11-1-0082.

\section{Supplemental information}

\appendix
\section*{Unconditional dynamics}
\subsection*{Master equation with tunable cavity}
In this section, we derive the master equation for the setup shown in Fig. 1 of the main text. We use the $(S,L,H)$ formalism \cite{GoughCommMathPhys2009} as this is a convenient method to derive the master equation for cascaded systems. A brief summary of the same is given below for self containment.

In this formalism, the quantum system is described by a triplet
\be
G\equiv(S,L,H),
\ee
where $S$ is the scattering matrix, $L$ is the vector of  coupling operators and $H$ is the Hamiltonian. In general for composing multiple systems, the following two products are defined. The concatenation product $\boxplus$ is used for composing subsystems into a system with	 stacked channels
\begin{eqnarray}
G_2 \boxplus G_1 &=& \left( \begin{pmatrix} S_2 & 0 \\ 0 & S_1 \end{pmatrix}, \begin{pmatrix} L_2 \\  L_1 \end{pmatrix}, H_2 + H_1 \right).  
\label{catproduct} 
\end{eqnarray}
The series product $\triangleleft$ of the triplets describes the feeding of output of one subsystem to another
\begin{eqnarray}
G_2 \triangleleft G_1 &=& \Bigg(S_2 S_1, S_2L_1 + L_2, \nonumber\\
&& H_1 + H_2 + \frac{1}{2i}\left(L_2^\dag S_2 L_1 - L_1^\dag S_2^\dag L_2 \right) \Bigg).
\label{seriesproduct} 
\end{eqnarray}
Using the above defined products, we can write down the $(S,L,H)$ triplet for the whole system 
\be
G = \left(S, \begin{pmatrix} L_1 \\ \vdots \\ L_n \end{pmatrix}, H\right),
\ee
from which we can extract the corresponding master equation as 
\be
\label{Eq:MEfromSLH}
\dot{\rho} = -i\comm{H}{\rho} + \sum_{i=1}^n \lind{L_i}\rho,
\ee
where the dissipation super-operator is given by $\lind{c} \rho = c\rho c^{\dagger }-\frac{1}{2}c^{\dagger }c\rho -\frac{1}{2} \rho c^{\dagger }c$.

The $(S,L,H)$ triplet for the case of a single transmon with 3 levels in front of a mirror is given by
\begin{eqnarray}
G_\text{tr} = \left(
{\mathbb{1}}_2,
\begin{pmatrix}
L_{01} \\
L_{12} 
\end{pmatrix} ,
H_\text{tr}
\right),
\end{eqnarray}
with the Hamiltonian in the rotating frame $H_\text{tr} = -\Delta_c \ket{0}\bra{0} + \Delta_p\ket{2}\bra{2}$, $L_{ij}=\sqrt{\Gamma_{ij}}\ket{i}\bra{j}$ where $\Gamma_{ij}$ is the relaxation rate for the $j\rightarrow i$ transition, $\Delta_c = \omega_{10} - \omega_c$ and $ \Delta_p = \omega_{21}-\omega_p$ are the detunings between the control (c)/probe (p) field and the corresponding transition frequencies of the transmon.

The $(S,L,H)$ triplets for the tunable cavity and the coherent probe in their corresponding rotating frames are
\be
G_{\text{cav}} = (1,\sqrt{\kappa(t)}a,0)
\ee
and
\be
G_{\alpha_p} = (1,\alpha_p,0), 
\ee
where $a$ ($a^\dag$) is the annihilation (creation) operator for the photons in the cavity, $\kappa(t)$ is the time dependent relaxation rate of the cavity and $\alpha_p$ is the complex amplitude of the probe.

Following the product rules in \Eqs{catproduct}{seriesproduct}, the $(S,L,H)$ triplet for the setup consisting of N cascaded transmons with a tunable cavity as the photon source and a coherent probe of strength $\alpha_p$ is
\begin{eqnarray}
G_\text{tot} &=&  G_\text{tr}^{(N)} \triangleleft \ldots \triangleleft G_\text{tr}^{(k)} \triangleleft \ldots \triangleleft G_\text{tr}^{(2)} \triangleleft G_\text{tr}^{(1)} \triangleleft
 \left(G_\text{cav} \boxplus G_{\alpha_p} \right) \nonumber\\
 && = {\left(
{\mathbb{1}}_2,
\begin{pmatrix}
\sqrt{\kappa(t)}a + \Lambda_{01}\\
\alpha_p + \Lambda_{12}
\end{pmatrix} ,
H_\text{tot}
\right),}
\end{eqnarray}
where
\begin{eqnarray}
H_\text{tot} &=& \sum_{j=1}^{N} H_\text{tr}^{(j)} + \frac{1}{2i}\sqrt{\kappa(t)}\left( \Lambda_{10} a - a^\dag \Lambda_{01} \right) \nonumber\\&&+ \frac{1}{2i}\left(\alpha_p \Lambda_{21} - \alpha_p^* \Lambda_{12} \right) \nonumber\\
&&+ \frac{1}{2i}\sum_{j=1}^N \left(\sum_{k=j+1}^N \left(L_{10}^{(k)} L_{01}^{(j)}  - L_{10}^{(j)} L_{01}^{(k)} \right)\right) \nonumber\\&& + \frac{1}{2i}\sum_{j=1}^N\left(\sum_{k=j+1}^N \left(L_{21}^{(k)} L_{12}^{(j)} - L_{21}^{(j)} L_{12}^{(k)} \right)\right)\nonumber\\
\end{eqnarray}
and the collective operators $\Lambda_{ij}=\sum\limits_{k=1}^{N} L_{ij}^{(k)}$. 
This gives the master equation
\begin{eqnarray}
\label{Eq_MECavity}
\dot{\rho} &=& -i\comm{H_\text{eff}}{\rho} + \sum_{j=1}^N\left(\lind{L_{01}^{(j)}} + \lind{L_{12}^{(j)}} \right)\rho \nonumber \\ &+& \kappa(t) \lind{a} \rho - \sqrt{\kappa(t)}\; \coup{a}{\Lambda_{01}} \rho \nonumber \\
&-& \sum_{j=1}^N \sum_{k=j+1}^N \bigg(\coup{L_{01}^{(j)}}{L_{01}^{(k)}} + \coup{L_{12}^{(j)}}{L_{12}^{(k)}}   \bigg)\rho \nonumber \\
&\equiv& \Lio_{\text{cav}}\rho, 
\end{eqnarray}
where we have defined a Liouvillian $\Lio_{\text{cav}}$ for shorthand. The effective Hamiltonian is $H_\text{eff} = \sum\limits_{k=1}^{N} H^{(k)}$ with $\text{H}^{(k)}= -\Delta^{(k)}_c\ket{0}\bra{0}^{(k)}+\Delta^{(k)}_p\ket{2}\bra{2}^{(k)} + \Omega_p(L^{(k)}_{12} + L^{(k)}_{21} )$. The normalization for the probe field has been chosen such that $\alpha_p=\Omega_p e^{i\pi/2}$, with $\Omega_p$ being a real number. To keep the notations compact, we have introduced a coupling super-operator $\coup{c_1}{c_2} \rho = \left[c_2^{\dagger }, c_1\rho \right]+\left[\rho~c_1^{\dagger },c_2\right]$.
\subsection*{Master equation with a Fock state input}
As an alternative approach to the setup with tunable cavity, we consider an arbitrary input Fock state photon interacting with the cascaded transmons. The formalism for this case is described in \citein{BaragiolaPRA2012}. As described there, the system dynamics is given by a set of coupled master equations for the generalized density matrices $\varrho_{m,n}$, where the indices $m$ and $n$  represent the Fock subspace and take values 0 or 1 for the case of a single photon input. We can proceed in the same way as in the previous section, use the $(S,L,H)$ formalism and derive the coupled set of master equations as
\begin{eqnarray}
\label{Eq_MasterFock}
\dot{\varrho}_{m,n} &=& -i\comm{H_\text{eff}}{\varrho_{m,n}} + \sum_{j=1}^N \Bigg(\lind{L_{01}^{(j)}} + \lind{L_{12}^{(j)}}\Bigg)\varrho_{m,n} \nonumber\\
&& - \sum_{j=1}^N \sum_{k=j+1}^N \Bigg(\coup{L_{01}^{(j)}}{L_{01}^{(k)}} +\coup{L_{12}^{(j)}}{L_{12}^{(k)}}\Bigg) \varrho_{m,n}  \nonumber\\
&& + \sqrt{m} \xi(t) \comm{\varrho_{m-1,n}}{\Lambda_{10}}+\sqrt{n} \xi^*(t)\comm{\Lambda_{01}}{\varrho_{m,n-1}}\nonumber\\
\end{eqnarray}
where $\xi(t)$ is the temporal shape of the photon wave packet.
Similar to the case with tunable cavity, we define a Liouvillian in this formalism and rewrite the master equation as
\bea
\label{Eq_MasterFock_Lio}
\dot{\varrho}_{m,n} &=& \Lio_{\text{Fock}}\;\varrho_{m,n} \nonumber\\ &+& \sqrt{m} \xi(t) \comm{\varrho_{m-1,n}}{\Lambda_{10}}+\sqrt{n} \xi^*(t)\comm{\Lambda_{01}}{\varrho_{m,n-1}}\nonumber\\
\eea
As explained in \citein{BaragiolaPRA2012}, the expectation values of the operators in this formalism are calculated using the top level density matrix. In the rest of the document, whenever we consider the Fock state formalism, we always have $\expec{A} = \tr{A\;\varrho_{1,1}}$ for any system operator $A$.

\begin{figure}[]
  \includegraphics[width=0.89\columnwidth]{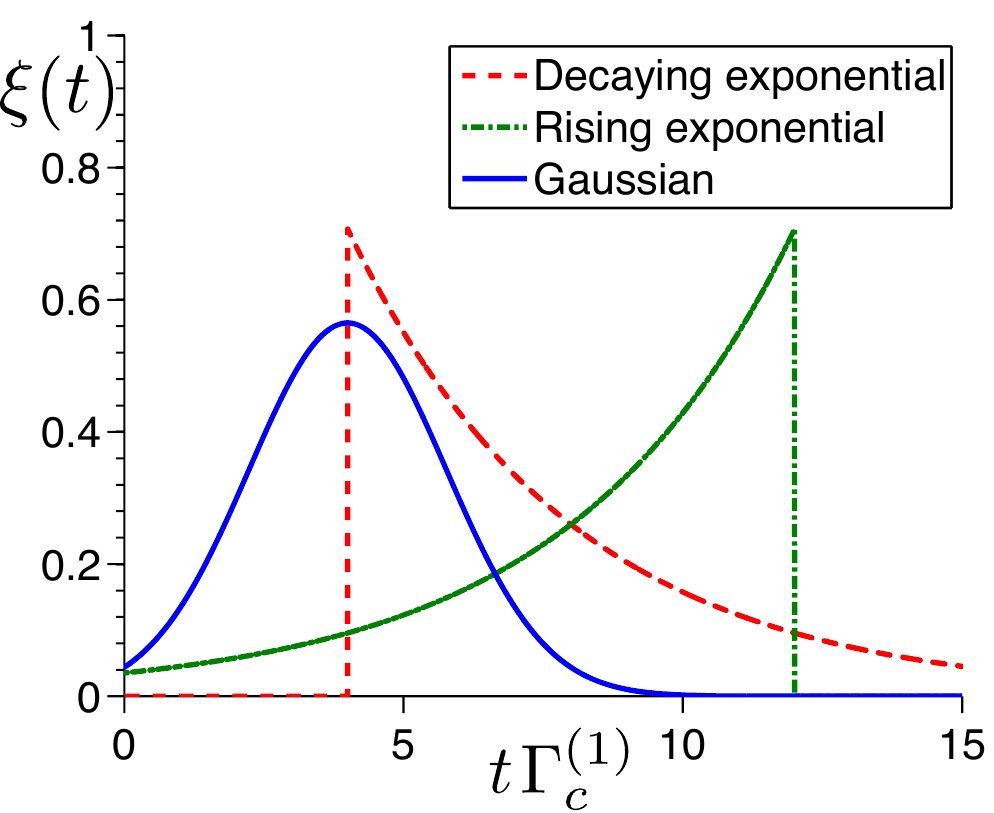}
\caption{Temporal shapes of the input control photon wave packets.}
\label{fig:input_photon_shape} 
\end{figure}

\section*{Measurement and conditional dynamics}
\subsection*{Stochastic master equation}
The conditional dynamics of the system due to homodyne detection of the output probe field is described by the stochastic master equation (SME). The SME can generally be considered as an unravelling of the average system dynamics described by the master \Eqs{Eq_MECavity}{Eq_MasterFock_Lio} \cite{CarmichaelQuantumOptics}.

In the case of tunable cavity as photon source, the SME is given by
\begin{eqnarray}
\label{Eq_SMECavity}
d\rho &=& \Lio_{\text{cav}}\rho dt  + \sqrt{\eta} \meas{\Lambda_{12}}{\rho} \; dW(t) ,
\end{eqnarray}
which is equation (2) in the main text. $\eta$ is the measurement efficiency, $dW(t)$ is a Wiener increment with $E[dW(t)]=0$ and $E[dW(t)^2]=dt$ where $E[\cdot]$ is the ensemble average. We have defined a measurement super-operator $\meas{c}{\rho} = (e^{i\phi}c\rho + e^{-i\phi}\rho c^{\dagger })-\expec{e^{i\phi}c+ e^{-i\phi} c ^{\dagger }}{\rho}$, that describes the back-action of the measurement on the evolution of the system.

Similarly, in the Fock state formalism the conditional dynamics is given by
\begin{align}
\label{Eq_SMEFock}
d\varrho_{m,n} &= \Lio_{\text{Fock}}\;\varrho_{m,n}dt \nonumber\\ 
&+ \sqrt{m} \xi(t) \comm{\varrho_{m-1,n}}{\Lambda_{10}}dt+\sqrt{n} \xi^*(t)\comm{\Lambda_{01}}{\varrho_{m,n-1}}dt \nonumber\\
&+ \sqrt{\eta} \meas{\Lambda_{12}}{\varrho_{m,n}} \; dW(t).
\end{align}
\subsection*{Signal}
From the solution of the SME, the homodyne current in the detector can be calculated as
\be
j (t) dt =  \sqrt{\eta}\expec{\hat{y}} dt + dW(t),
\ee
where $\hat{y} = e^{i\phi}\Lambda_{12}+e^{-i \phi}\Lambda_{21}$, with the phase of the local oscillator $\phi$ specifying the quadrature of measurement. All the information that can be extracted from the system is contained in $j(t)$. To convert this into a binary outcome, we define the signal of detection S as the integrated current during the measurement time $t_m = t_{f}-t_{i}$ 
\[S = \int_{t_{i}}^{t_{f}} j(t) dt.\]

The signal to noise ratio (SNR) is then defined as
\be
\label{SNR_def}
\text{SNR}=\frac{E[S_1]-E[S_0]}{\sqrt{\text{Var}[S_1]+\text{Var}[S_0]},} 
\ee
where $S_{0/1}$ is the signal with 0 or 1 photon in the control field and $\text{Var}[X]=E[X^2]-E[X]^2$ is the variance.

In the case of stochastic master equations (\Eqs{Eq_SMECavity}{Eq_SMEFock}), the evolution of the system is simulated for 2000 trajectories with and without a control photon. The average and the variance are calculated from the distribution of the signal.
 
The SNR can also be calculated from the unconditional dynamics (\Eqs{Eq_MECavity}{Eq_MasterFock_Lio})\cite{QuantMeasAndContrl}. The average and variance in  this case are
\[E[S_0]=0 \;;\;E[S_0^2]=t_m\]
\[ E[S_1]=\sqrt{\eta}\int_{t_{i}}^{t_{f}} \expec{\hat{y}} dt\]
\be 
\label{Eq_ttc}
E[S_1^2] =  \int_{t_{i}}^{t_{f}} dt_1 \int_{t_{i}}^{t_{f}} dt_2 \;E[j(t_1) j(t_2)].
\ee
In the case of tunable cavity as photon source, the two time correlation function can be evaluated using the quantum regression theorem \cite{Lax66} as
\begin{widetext}
\bea
\label{ttc_cav_def}
E[j(t_1) j(t_2)] &=& \Theta(t_2-t_1)\bigg(\eta \tr{(e^{i\phi}\Lambda_{12} + e^{-i\phi} \Lambda_{21})\mathcal{T}(t_2-t_1)(e^{i\phi}\Lambda_{12} \rho(t_1) + e^{-i\phi} \rho(t_1) \Lambda_{21})} +\delta(t_2-t_1)\bigg) \nonumber\\ &+& \Theta(t_1-t_2)\bigg(\eta \tr{(e^{i\phi}\Lambda_{12} +e^{-i\phi} \Lambda_{21})\mathcal{T}(t_1-t_2)(e^{i\phi}\Lambda_{12} \rho(t_2) +e^{-i\phi} \rho(t_2) \Lambda_{21})} +\delta(t_1-t_2)\bigg)\nonumber\\ 
\eea
\end{widetext}
where $\mathcal{T}(t_2-t_1) \mathbb{Y}(t_1) = \mathbb{Y}(t_2)$ with $\mathbb{Y}(t)=e^{i\phi}\Lambda_{12} \rho(t) + e^{-i\phi} \rho(t) \Lambda_{21}$. The time evolution operator $\mathcal{T}(t_2-t_1)$ is evaluated by solving $\dot{\mathbb{Y}} = \Lio_{\text{cav}}\mathbb{Y}$. By definition, the step function $\Theta(t)=0$ for $t<0$ and 1 otherwise.

The procedure to calculate two time correlations in the Fock state formalism is presented in the next section. 
\subsection*{Quantum regression theorem for Fock state input}
\label{sec_QRT}
From the formalism presented in \citein{GJN12,BaragiolaPRA2012}, it is not immediately clear  how to apply the quantum regression theorem for Fock-state input fields, although the authors of \citein{GheZol99} hinted at the method for a single photon.  Here, we present a brief derivation with details to be presented elsewhere.  For system operators $A$ and $B$ we wish to find 
		\begin{equation}
			\expec{A(t)B(t+\tau)} = \Tr_{\rm sys+field}[A(t) B(t+\tau) \rho_{\rm tot}],
		\end{equation}
where $\rho_{\rm tot}$ is the total density matrix of system + field.

The unitary evolution of the system + field during an infinitesimal time interval is given by\cite{GoughIEEE2009}
\be
\label{FockUnitary}
U(t+dt,t)=\mathbb{1}-\frac{1}{2}L^\dagger L dt - L^\dagger S dB_t + (S-\mathbb{1})d\Lambda
\ee
where the quantum noise increments satisfy the It$\bar{\rm o}$ table
\be
\begin{array}{c|c}

dB dB^\dagger = dt  & dB d\Lambda = dB \\ 
\hline
d\Lambda dB^\dagger = dB^\dagger & d\Lambda d\Lambda = d\Lambda \\ 

\end{array} 
\ee
with all other products vanishing. Further details on these noise increments is given in  \citein{BaragiolaPRA2012}. The composition property of the unitary evolution operator \big($U(t,t_0) = U(t,t') U(t',t_0)$ for $t \geq t' \geq t_0$\big) and the cyclic property of trace, allows us to write
	 	\begin{align}
	 	\label{Eq::TwoTimeCorrelation}
		&\expec{A(t)B(t+\tau)}=&& \nonumber\\ &\Tr_{\rm sys} \Big[  B \, \Tr_{\rm field} \big[ U(t+ \tau,t)  \rho_{\rm tot}(t) A U\dg(t+ \tau,t)  \big] \Big].\nonumber\\
	\end{align}
For a system interacting with an $N$-photon Fock state, the initial state of the system and field is
	\begin{equation} \label{Eq::FockInitialState}
		\rho_{\rm tot}(t_0) = \rho_{\rm sys} \otimes \op{N_\xi}{N_\xi}.
	\end{equation}
		
Through the unitary interaction, governed by \Eq{FockUnitary}, it can be shown that the physical reduced density matrix for the system couples to a set of \emph{generalized density matrices} (GDMs)
	\begin{align} \label{Eq::GDMs}
		\varrho_{m,n}(t) = \Tr_{\rm field} \big[ U(t,t_0 )\rho_{\rm sys} \otimes \op{m_\xi}{n_\xi} U\dg(t,t_0 ) \big],
	\end{align}
where the indices $\{m,n\}$ take on integer values between 0 and the number of photons in the input field $N$.  To extract the reduced system density operator at some later time $t$, one must propagate the set of GDMs.    

	Similarly, if we plug \Eq{Eq::FockInitialState} into \Eq{Eq::TwoTimeCorrelation}, we can define a two-time system operator,
	\begin{align}
		\mathbb{T}_{N,N}(t+\tau, t)\equiv& \Tr_{\rm field} \bigg[ U(t+ \tau,t) U(t,0)  \rho_{\rm sys} \nonumber \\ &\otimes \op{N_\xi}{N_\xi} U\dg(t,0) A U\dg(t+ \tau,t)  \bigg]
	\end{align}
subject to the boundary condition $\mathbb{T}_{N,N}(t, t) = \varrho_{N,N}(t) A.$  Just as for the Fock-state master equations, the physical two-time operator $\mathbb{T}_{N,N}(t', t)$ couples via the unitary dynamics to auxiliary two-time operators $\mathbb{T}_{m,n}(t', t)$ defined  as
	\begin{align}
		\mathbb{T}_{m,n}(t', t)\equiv& \Tr_{\rm field} \bigg[ U(t',t) U(t,0)  \rho_{\rm sys} \nonumber \\ &\otimes \op{m_\xi}{n_\xi} U\dg(t,0) A U\dg(t',t)  \bigg].
	\end{align}
Using the explicit form for the time evolution operator (\Eq{FockUnitary}), it can be shown that the two-time operators satisfy a set of coupled differential equations
\begin{widetext}
	\begin{align} \label{Eq::TwoTimeEOMs}
		\frac{d}{dt'}\mathbb{T}_{m,n}(t', t)  = &   \mathcal{D}[L]  \mathbb{T}_{m,n}(t', t)  +  \sqrt{m} \xi(t') \big[S \mathbb{T}_{m-1,n}(t', t), L\dg \big] + dt \sqrt{n} \xi^*(t') \big[L, \mathbb{T}_{m,n-1}(t', t) S\dg \big]  \\
			& +   \sqrt{mn} |\xi(t')|^2 \left( S \mathbb{T}_{m-1,n-1}(t', t) S\dg - \mathbb{T}_{m-1,n-1}(t', t) \right) \nn
	\end{align}
\end{widetext}
subject to the boundary conditions $\mathbb{T}_{m,n}(t, t) = \varrho_{m,n}(t) A.$  To calculate two-time correlation functions, follow this recipe.  First, calculate all the GDMs at time $t$.  Second,  evolve all of the two-time operators $\mathbb{T}_{m,n}(t',t)$ from time $t$ to $t + \tau$ using \Eq{Eq::TwoTimeEOMs}.  Finally, take the trace of the physical two-time operator with system operator $B$,
	\begin{align}
		\expec{A(t)B(t+\tau)} & = \Tr_{\rm sys} \big[ B \mathbb{T}_{N,N}(t+\tau, t) \big].
	\end{align}

In particular, for calculating the two time correlation in \Eq{Eq_ttc}, we define a set of operators 
\[\mathbb{Y}_{m,n}(t,t)=e^{i\phi}\Lambda_{12} \varrho_{m,n}(t) + e^{-i\phi} \varrho_{m,n}(t) \Lambda_{21}\]
and evolve these to later time $t+\tau$ using \Eq{Eq::TwoTimeEOMs} which in our case becomes,
\bea
\frac{d}{dt'}\mathbb{Y}_{m,n}(t',t) &=& \Lio_{\text{Fock}}\;\mathbb{Y}_{m,n}(t',t) \nonumber\\ &+& \sqrt{m} \xi(t') \comm{\mathbb{Y}_{m-1,n}(t',t)}{\Lambda_{10}}\nonumber\\ &+&\sqrt{n} \xi^*(t')\comm{\Lambda_{01}}{\mathbb{Y}_{m,n-1}(t',t)}.\nonumber\\
\eea 
Using the solution from the above equation, we calculate the two time correlation as
\bea
\label{ttc_fock_def}
E[j(t_1) j(t_2)] &=& \Theta(t_2-t_1)\big(\eta \Tr \left[\hat{y}\;\mathbb{Y}_{1,1}(t_2,t_1)\right] +\delta(t_2-t_1)\big) \nonumber\\ &+& \Theta(t_1-t_2)\big(\eta \Tr \left[\hat{y}\;\mathbb{Y}_{1,1}(t_1,t_2)\right] +\delta(t_1-t_2)\big).\nonumber\\ 
\eea
\subsection*{Output photon flux}
Fig. 2d in the main text and \Fig{fig:SP} show the output field quantities from the 0-1 transition of the transmon. These are calculated using the Fock state formalism following  \citein{BaragiolaPRA2012}. The mean output photon flux from the 0-1 transition at the end of the transmon chain is
\begin{eqnarray}
\label{Eq_flux}
|\xi_{out}(t)|^2&=&\mathbb{E}_{1,1}[\Lambda_{10} \Lambda_{01}]+\xi^*(t)\mathbb{E}_{0,1}[\Lambda_{01}]+\xi(t)\mathbb{E}_{1,0}[\Lambda_{10}]  \nonumber\\
&& +|\xi(t)|^2,\nonumber\\
\end{eqnarray}
where $\mathbb{E}_{m,n}[O]=\tr{\varrho_{m,n}^\dagger O}$. The integrated flux upto time T is 
\be E = \int_0^T |\xi_{out}(t)|^2 dt .\ee
\section*{SNR dependence on system imperfections}
\subsection*{Dependence on transmon and signal parameters}
The results in Fig. 2 of the main text were obtained for a specific set of parameter values. In this section, we show the dependence of SNR on these values for a given number of transmons $N=6$. Since the total number of parameters that can be tuned is large ($\sim 25$ for $N=6$), we show only few relevant results here.

\begin{figure}[]
  \includegraphics[width=\columnwidth]{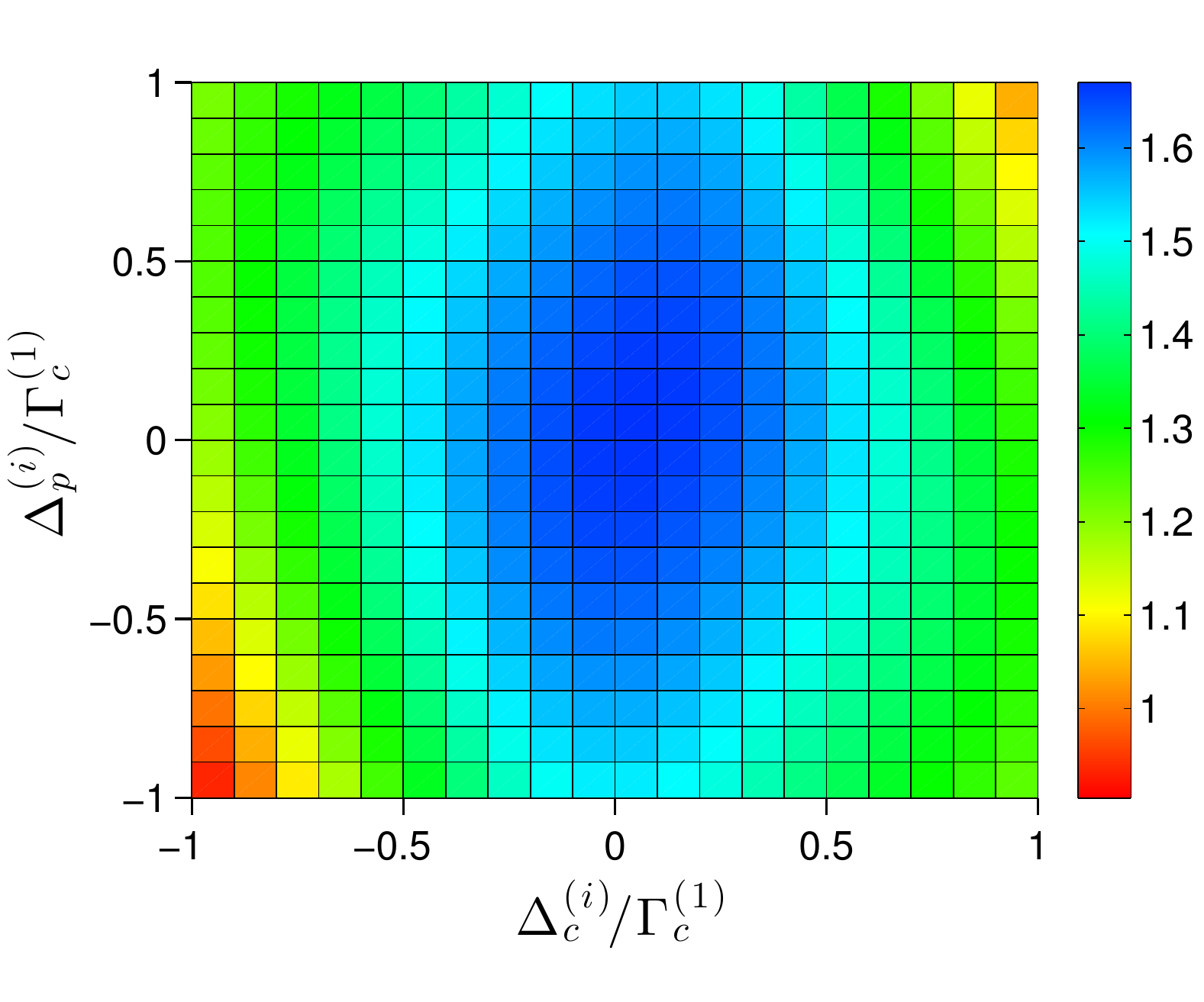}
\caption{SNR for photon detection with $N=6$ transmons in the chain as a function of the detuning of the control and probe of all the transmons. The input control photon is of Gaussian temporal shape. The values of other parameters are the same as in Table I of the main text.}
\label{fig:detuning} 
\end{figure}

\begin{figure}[]
  \includegraphics[width=\columnwidth]{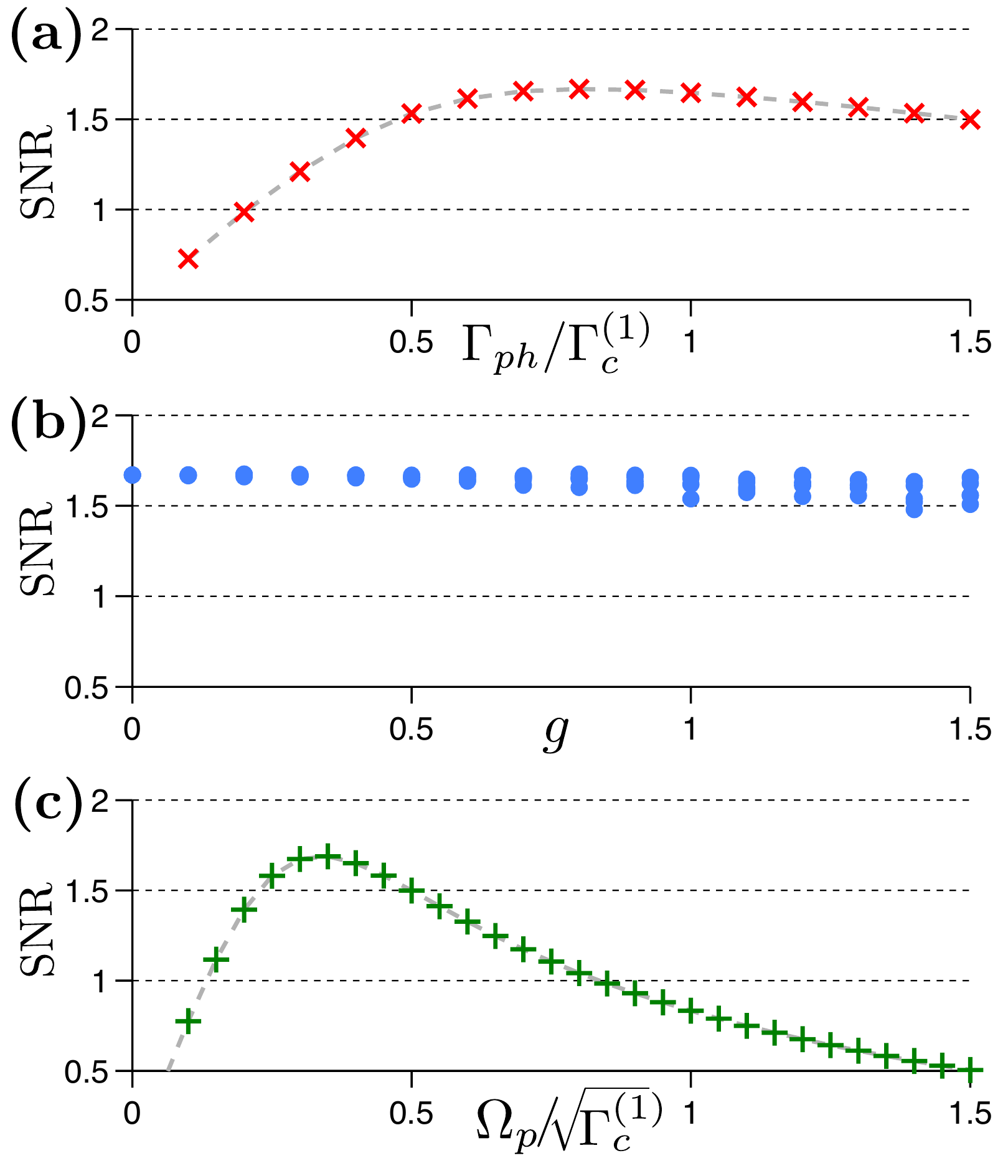}
\caption{a) SNR as a function of the width of the input photon. b) Effect of coupling on SNR as a function of parameter g defined in the text for 5 random data sets. c)Effect of probe strength on the SNR. In all of the above cases, we have $N=6$ transmons in the chain  and the input control photon is of Gaussian temporal shape. The values of other parameters are the same as in Table I of the main text.}
\label{fig:Gamma} 
\end{figure}

In \Fig{fig:detuning}, we plot the SNR when the detuning of the control $\Delta_c$  and probe $\Delta_p$ for all the transmons are changed. \Fig{fig:Gamma}a gives the dependence of SNR on the width of the input control photon. In order to illustrate the effect of changing the coupling of 0-1 transition on SNR, we define a new parameter g that describes the variation from the values in Table I. The coupling of the control photon to the 0-1 transition for the 6 transmons is then given by 
\bea
\label{Eq_Gammac_values}
\Gamma_c &= \big\{& 1 ; 1.9+\delta g^{(2)} ; 2.2+\delta g^{(3)} ; \\ \nonumber && 2.5+\delta g^{(4)} ; 2.4+\delta g^{(5)} ; 2.5+\delta g^{(6)}\big\},\\\nonumber
\eea where $\delta g^{(i)}$ is a uniformly distributed random number between -g and g. 
\Fig{fig:Gamma}b shows the variation of SNR with this parameter for 5 different sets of random samples. In \Fig{fig:Gamma}c, the dependence of SNR on the probe strength is shown. 

From these figures, it is clear that the proposed setup is robust enough to allow for imperfections that might occur in experimental realizations.


\subsection*{Dependence on pure dephasing}
\begin{figure}[]
  \includegraphics[width= 0.9\columnwidth]{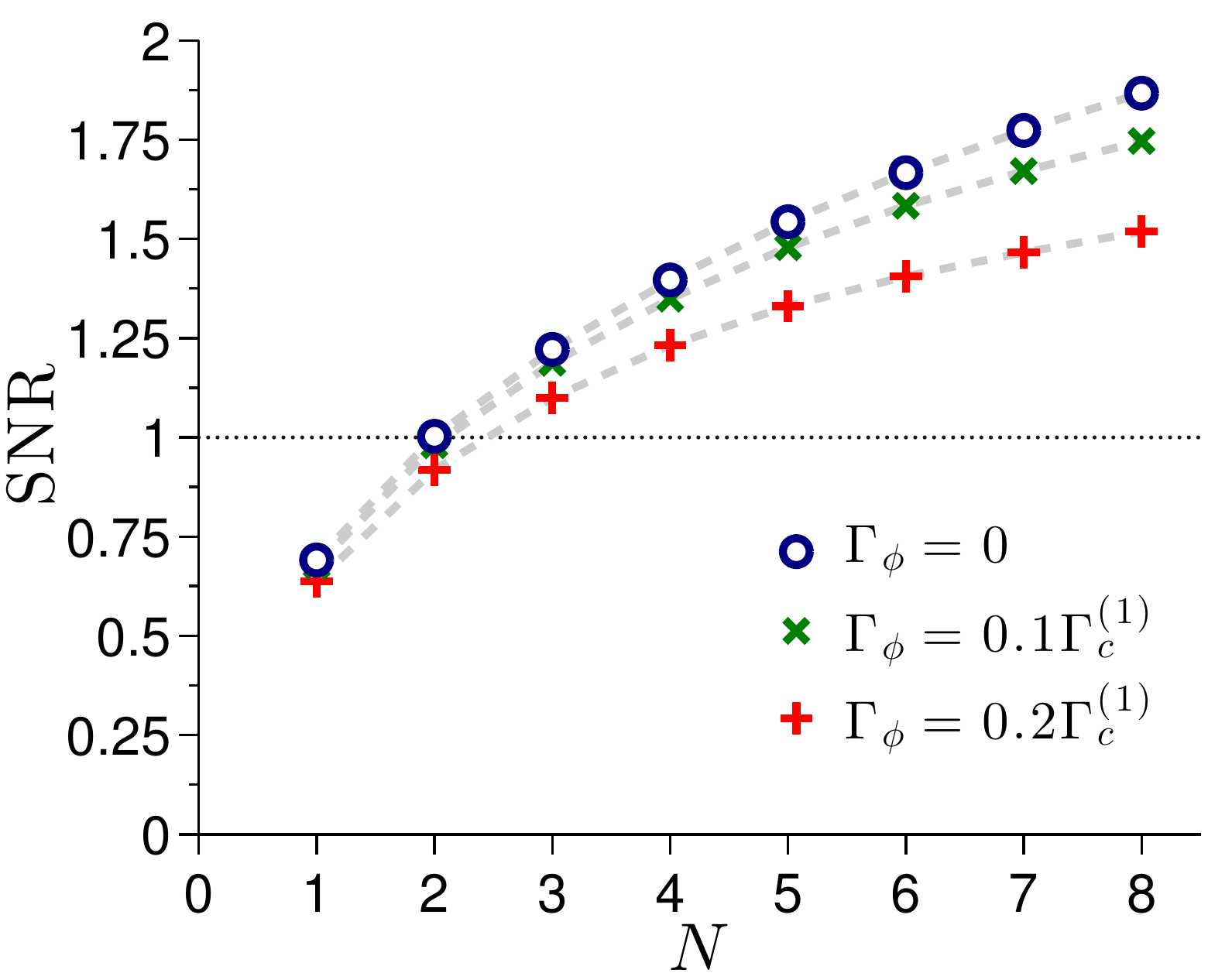} 
     \caption{SNR as a function of number of transmons for different values of dephasing for a Gaussian input photon. The value of other parameters are given in Table I of the main text.}
     \label{fig:dephasing}
\end{figure}

To include the effect of dephasing of the individual transmons on SNR, the master equations (\Eqs{Eq_MECavity}{Eq_MasterFock_Lio}) are modified as
\begin{eqnarray}
\label{Eq_cavity_deph}
\dot{\rho} &=& \Lio_{\text{cav}}\rho + \sum_{j=1}^N \frac{\Gamma_{\phi}^{(j)}}{2}\lind{2\sigma_{11}^{(j)} +4\sigma_{22}^{(j)}}\rho 
\end{eqnarray}
and
\begin{eqnarray}
\label{Eq_MasterFock_deph}
\dot{\varrho}_{m,n} &=& \Lio_{\text{Fock}}\varrho_{m,n} \nonumber\\ 
&+& \sqrt{m} \xi(t) \comm{\varrho_{m-1,n}}{\Lambda_{10}}+\sqrt{n} \xi^*(t)\comm{\Lambda_{01}}{\varrho_{m,n-1}} \nonumber \\ &+& \sum_{j=1}^N \frac{\Gamma_{\phi}^{(j)}}{2}\lind{2\sigma_{11}^{(j)} +4\sigma_{22}^{(j)}}\varrho_{m,n}.
\end{eqnarray}
\Fig{fig:dephasing} shows the effect of dephasing on the scaling of SNR with number of transmons. We see that for experimentally relevant values of dephasing\cite{IoChunArxiv}, the impact on SNR is not too significant.

\section*{Shape preserving mode}
\begin{figure}[]
 \includegraphics[width= \columnwidth]{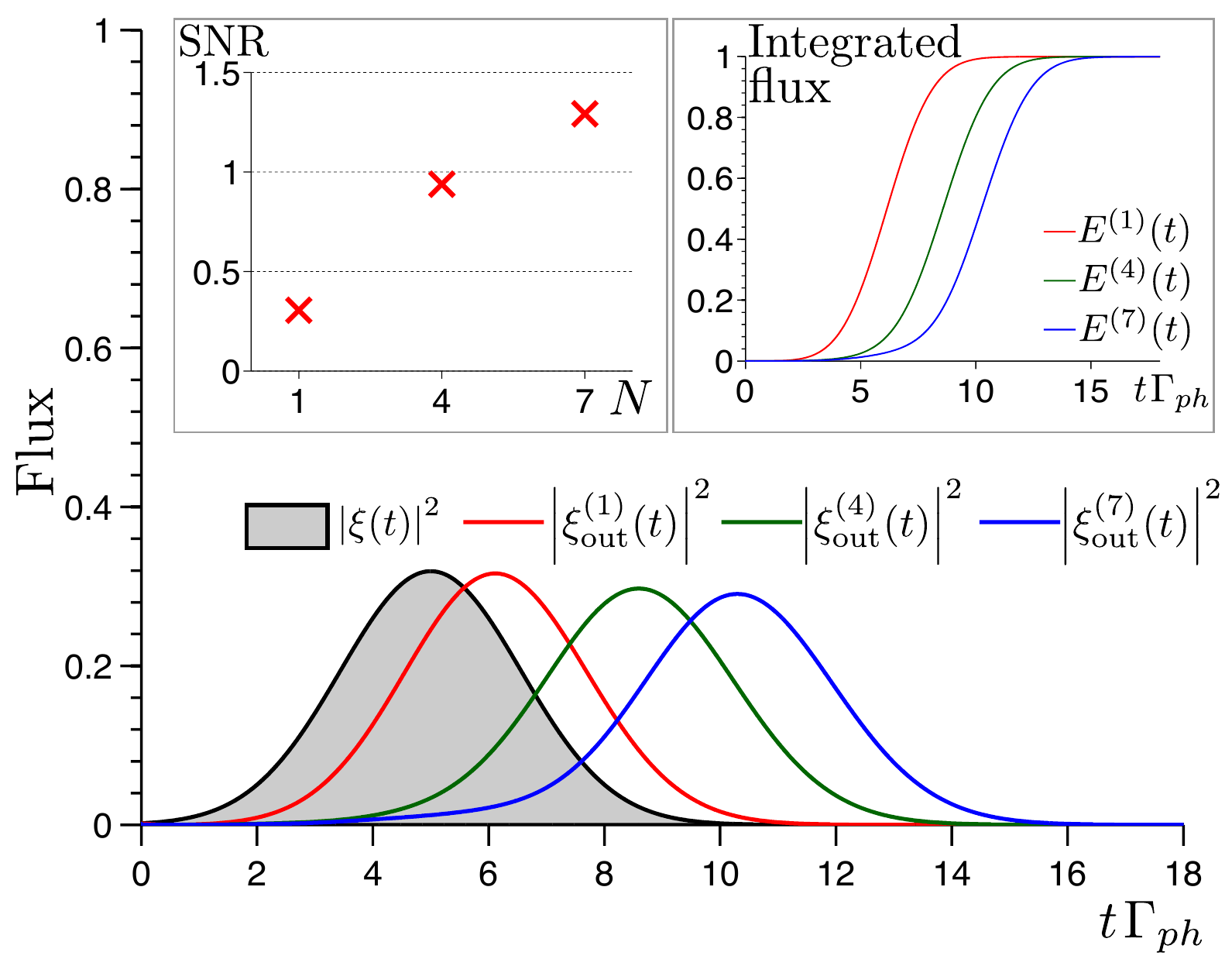} 
     \caption{
     SNR with the detector parameters optimized to preserve the temporal shape of the incoming photon flux (grey shaded). The SNR is slightly lower than the result in Fig. 2 of the main text. The figure shows the output photon flux, the integrated flux and the corresponding SNR  after $N=\{1,4,7\}$ transmons. In addition to being QND in photon number, the detector can be tuned to minimally distort the shape of the  photon flux. The parameters were tuned for a given number of transmons, to preserve the shape while still giving a significant homodyne current.}
     \label{fig:SP}
 \end{figure}

Although the proposed detector is nondestructive in the number of photons in the control field, the interaction with the detector causes the photon envelope to become distorted (refer Fig. 2d in the main text). We note that since the energy levels and coupling strengths of each transmon are tunable, there is enough flexibility in our proposal to minimize the wave packet distortion while still being able to detect it. In \Fig{fig:SP} we show the SNR as a function of transmon number with parameters optimized to preserve photon shape.  In this parameter regime the SNR is slightly lower than that obtained in Fig. 2 of the main text, with the number of transmons $N = 5$ needed to achieve a SNR above unity. The shape of the output photon flux is however more similar to the input state, as can be seen in the \Fig{fig:SP}, where we plot the integrated photon flux along with the shape of the incoming and transmitted photon flux. Although this regime does require more transmons, the overhead is moderate indicating that it is possible to work our proposal as a detector which is QND in photon number and, to a large extent, also in temporal shape.\\
\\
\section*{Effect of losses in circulators}
\subsection*{Modeling imperfections with beam splitters}

\begin{figure*}[]
 \includegraphics[width= 0.9\textwidth]{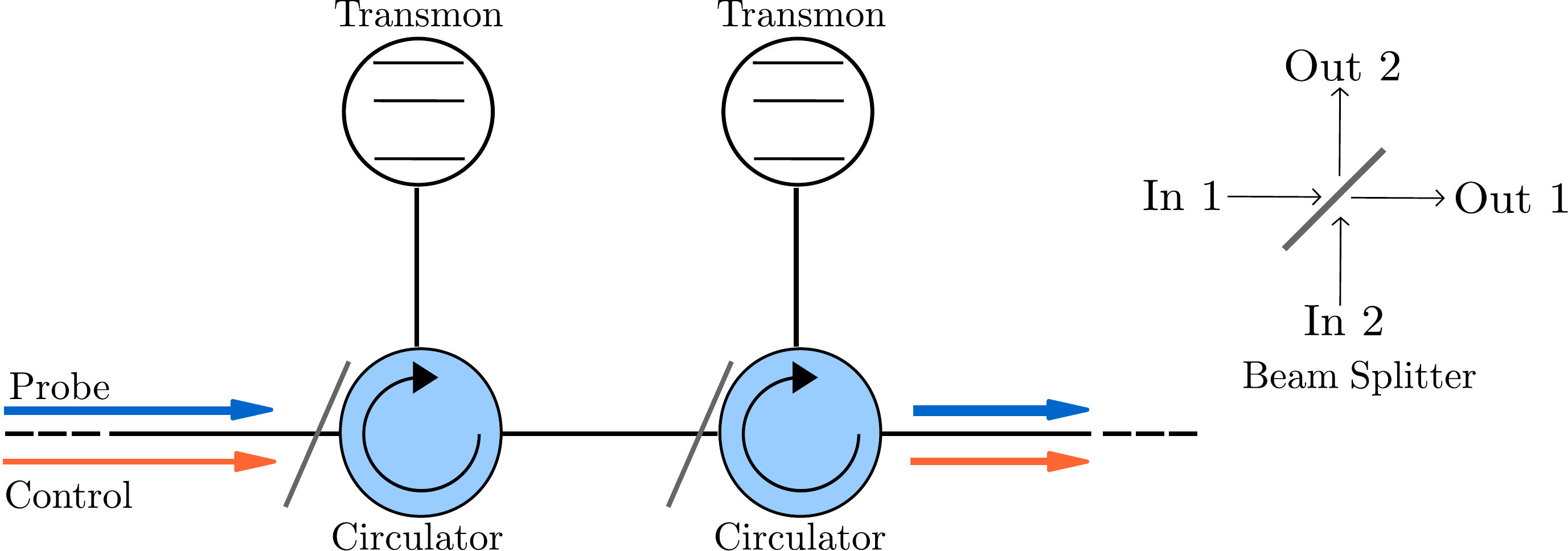} 
     \caption{A schematic depiction of the photon detection with beamsplitters added, and also a sketch of a single beamsplitter. The control photon and the coherent probe enter the setup from the left.}
     \label{fig:TransmonsWithBS}
 \end{figure*}
 
To model the imperfections in the circulators, we place fictitious beam splitters in front of every circulator and think of the circulators as perfect (refer \Fig{fig:TransmonsWithBS}). The beam splitters then give a small loss, which accounts for attenuation loss, reflection at the circulators and routing of the signal to the wrong port at the circulator (isolation loss). Since isolation makes the photon miss a transmon in the chain, the efficiency is decreased but the photon can still interact with transmons further down the chain. By modeling this type of error as an attenuation loss, we achieve a lower bound on the measurement efficiency. We don't consider the possibility of back-scattered signal interacting with the transmons, since it's a second order effect (it requires first reflection, then routing to the wrong port). Reflections from circulators going back immediately to a transmon just gives renormalization of the transmon parameters (if the travel time is short) or looks like loss (if the travel time is long). 

The $(S,L,H)$ component for a beam splitter with a transmission amplitude $t$ and reflection amplitude $r$ is
\bea
G_\text{BS} = \left(
\begin{pmatrix}
t & r \\
-r & t
\end{pmatrix}
,
\begin{pmatrix}
0 \\
0
\end{pmatrix}
,
0
\right),
\eea
where $t=\sqrt{1-r^2}$ ensures the unitarity of the scattering matrix. The power loss from each circulator is given by $P_{loss} = |r|^2$.

It must be noted that each beam splitter adds an additional channel into which the signal is lost, which is included in the $(S,L,H)$ product via the triplets defined as
\be
I_{d} = (\mathbb{1}_d,0,0), 
\ee
where $\mathbb{1}_d$ is the $d$-dimensional identity matrix. 

Following the product rules in \Eqs{catproduct}{seriesproduct}, we can derive the  $(S,L,H)$ triplet for the setup consisting of cascaded transmons including the beam splitters with a tunable cavity as the photon source. We do this by following an iterative approach starting from $N=1$ as
\begin{eqnarray}
G_\text{tot,1} &=& \left(G_\text{tr} \boxplus I_1 \right) \triangleleft G_\text{BS} \triangleleft
 \left(G_\text{cav} \boxplus G_{\alpha_p} \right) \nn \\
G_\text{tot,2} &=& \left(G_\text{tr}^{(2)} \boxplus I_2 \right) \triangleleft \left( G_\text{BS}^{(2)} \triangleleft \left( G_\text{tot,1} \boxplus I_1\right) \right) \nn\\
G_\text{tot,3} &=& \left(G_\text{tr}^{(3)} \boxplus I_3 \right) \triangleleft \left( G_\text{BS}^{(3)} \triangleleft \left( G_\text{tot,2} \boxplus I_1\right) \right) \nn
\end{eqnarray}
and so on, to get the full $(S,L,H)$ triplet $G_\text{tot,N}$.  From the full triplet, we can write the master equation using \Eq{Eq:MEfromSLH} as 
\begin{eqnarray}
\label{Eq_MECavity_circloss}
\dot{\rho} &=& -i\comm{H_\text{eff}}{\rho} + \sum_{j=1}^N\left(\lind{L_{01}^{(j)}} + \lind{L_{12}^{(j)}} \right)\rho \nonumber \\ &+& \kappa(t) \lind{a} \rho - \sum_{j=1}^N t^j \sqrt{\kappa(t)}\; \coup{a}{L_{01}^{(j)}} \rho \nonumber \\
&-& \sum_{j=1}^N \sum_{k=j+1}^N \bigg(\coup{L_{01}^{(j)}}{L_{01}^{(k)}} + \coup{L_{12}^{(j)}}{L_{12}^{(k)}}   \bigg)\rho , \nonumber \\ 
\end{eqnarray}
where we have also used the fact that $r^2+t^2=1$. The effective Hamiltonian in the above master equation is $H_\text{eff} = \sum\limits_{k=1}^{N} H^{(k)}$ with $\text{H}^{(k)}= -\Delta^{(k)}_c\ket{0}\bra{0}^{(k)}+\Delta^{(k)}_p\ket{2}\bra{2}^{(k)} + t^{(k)} \Omega_p(L^{(k)}_{12} + L^{(k)}_{21} )$. 

\begin{figure}[]
  \includegraphics[width=\columnwidth]{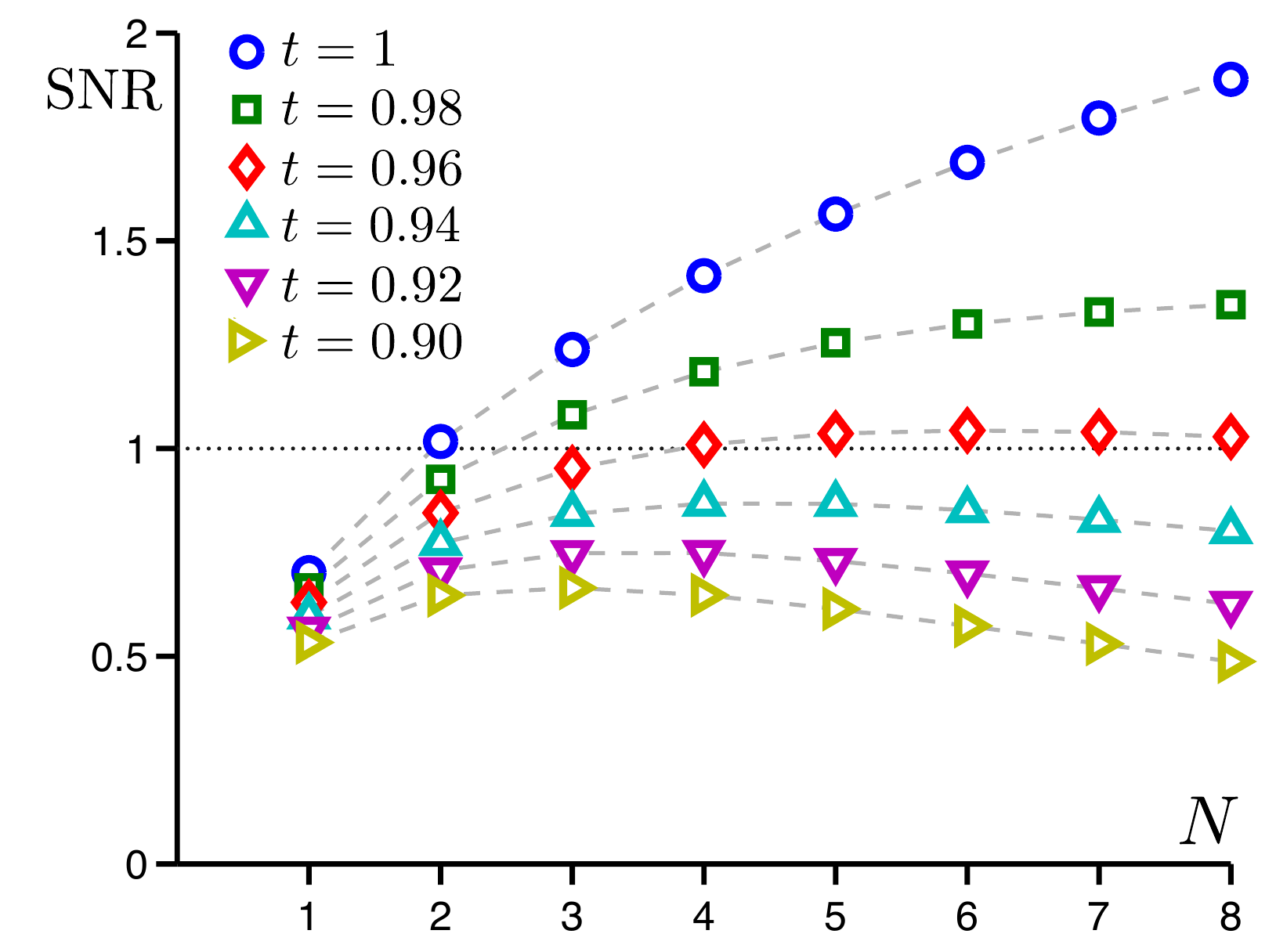}
\caption{Effect of losses in circulators on SNR for different values of transmission amplitudes $t$. The input control photon is Gaussian and the parameter values are as in Table I of the main text.}
\label{fig:circ_loss} 
\end{figure}

The homodyne current in this case is given by 
\be
j(t) dt = \sqrt{\eta}\expec{y} + dW(t),
\ee
where
\be
y = \sum_{j=1}^N t^{N-j} \left(e^{i\phi} L_{12}^{(j)} + e^{-i\phi} L_{21}^{(j)} \right).
\ee
The SNR is then calculated as in the previous sections. The result is shown in \Fig{fig:circ_loss} for different values of transmission amplitudes.
\section*{Measurement with a linear filter}
In this section, we check if the SNR of photon detection can be improved by using a better filter for integration. In this case, we define the  signal of detection S as the integrated current during the measurement time $t_m$ 
\[S = \int_{0}^{t_m} f(t) j(t) dt,\]
where $f(t)$ is a filter function over which the integration is done.
 
The SNR is calculated from the unconditional dynamics using \Eq{SNR_def}. The average and variance are given by
\[E[S_0]=0 \;;\;E[S_0^2]=\int_{0}^{t_m} (f(t))^2 dt\]
\[ E[S_1]=\sqrt{\eta}\int_{0}^{t_m} f(t) \expec{\hat{y}} dt\]
\[E[S_1^2] =  \int_{0}^{t_m} dt_1 \int_{0}^{t_m} dt_2 \;f(t_1) f(t_2) E[j(t_1) j(t_2)],\]
where the two-time correlations are calculated using \Eq{ttc_cav_def} or \Eq{ttc_fock_def}.

\begin{figure}[]
  \includegraphics[width=\columnwidth]{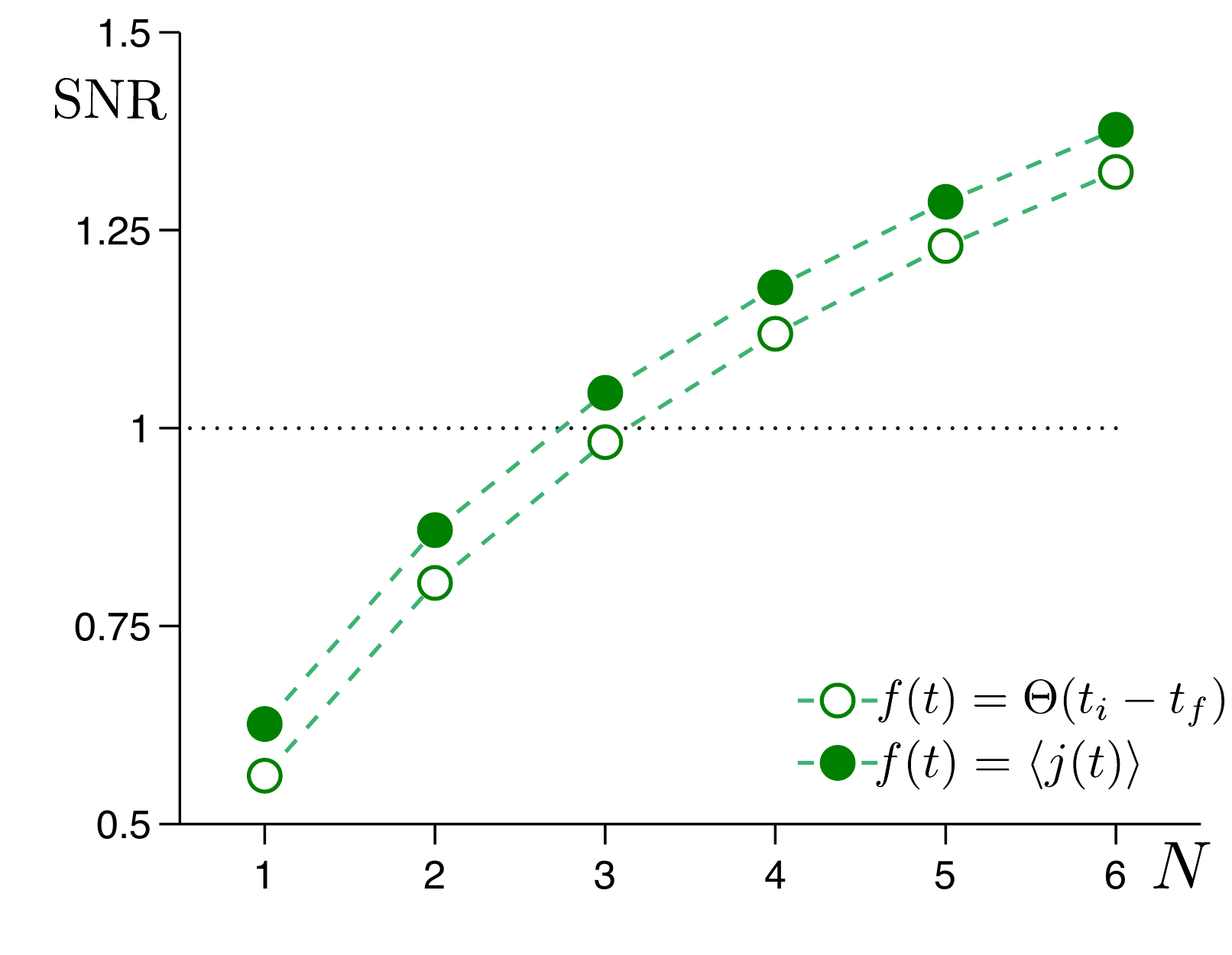}
\caption{SNR for measurements done with a linear filter. The input control photon is a rising exponential. The parameter values are as given in Table I of the main text. For $f(t) = \Theta(t_i-t_f),$ the values of $t_i$ and $t_f$ were optimized numerically. For the filter function, $f(t)=\expec{j(t)},$ the measurement is done from $t=0$ till the homodyne current becomes zero, so as to account for all the signal.}
\label{fig:filters} 
\end{figure}

The SNR in the main text and previous sections were obtained using a filter function $f(t) = \Theta(t_i - t_f)$. Here, we compare those results with the SNR obtained using $f(t) = \expec{j(t)}.$ The difference is shown in \Fig{fig:filters} for a input control photon that is a rising exponential. As evident from the figure, the detection efficiency increases with better filtering but the effect is marginal for this simple filter. We also note that the detection scheme could be further improved by using non-linear filters \cite{GambettaPRA2007}.

\end{document}